\documentclass{vgtc}                          

\ifpdf
  \pdfoutput=1\relax                   
  \usepackage{graphicx}                
  \DeclareGraphicsExtensions{.pdf,.png,.jpg,.jpeg} 
\else
  \usepackage{graphicx}                
  \DeclareGraphicsExtensions{.eps}     
\fi%

\graphicspath{{figures/}{pictures/}{images/}{./}} 

\usepackage{microtype}                 
\PassOptionsToPackage{warn}{textcomp}  
\usepackage{textcomp}                  
\usepackage{mathptmx}                  
\usepackage{times}                     
\usepackage{cite}                      
\usepackage{tabu}                      
\usepackage{booktabs}                  
\usepackage{amssymb}
\usepackage[usenames, dvipsnames]{color}
\usepackage{url}
\onlineid{1098}

\vgtccategory{Research}

\vgtcinsertpkg

\title{Segue: Overviewing Evolution Patterns of Egocentric Networks \\ by Interactive Construction of Spatial Layouts}

\author{
Po-Ming Law \thanks{e-mail: pmlaw@gatech.edu}\\ \scriptsize Georgia Institute of Technology \and Yanhong Wu \thanks{e-mail: yanwu@visa.com}\\ \scriptsize Visa Research \and Rahul C. Basole \thanks{e-mail: basole@gatech.edu}\\ \scriptsize Georgia Institute of Technology}

\abstract{Getting the overall picture of how a large number of ego-networks evolve is a common yet challenging task. Existing techniques often require analysts to inspect the evolution patterns of ego-networks one after another. In this study, we explore an approach that allows analysts to interactively create spatial layouts in which each dot is a dynamic ego-network. These spatial layouts provide overviews of the evolution patterns of ego-networks, thereby revealing different global patterns such as trends, clusters and outliers in evolution patterns. To let analysts \textit{interactively} construct \textit{interpretable} spatial layouts, we propose a data transformation pipeline, with which analysts can adjust the spatial layouts and convert dynamic ego-networks into event sequences to aid interpretations of the spatial positions. Based on this transformation pipeline, we developed Segue, a visual analysis system that supports thorough exploration of the evolution patterns of ego-networks. Through two usage scenarios, we demonstrate how analysts can gain insights into the overall evolution patterns of a large collection of ego-networks by interactively creating different spatial layouts.}

\CCScatlist{
  \CCScatTwelve{Human-centered computing}{Visu\-al\-iza\-tion}{Visu\-al\-iza\-tion techniques}{Graph drawings}
}

\teaser{
  \includegraphics[width=\linewidth]{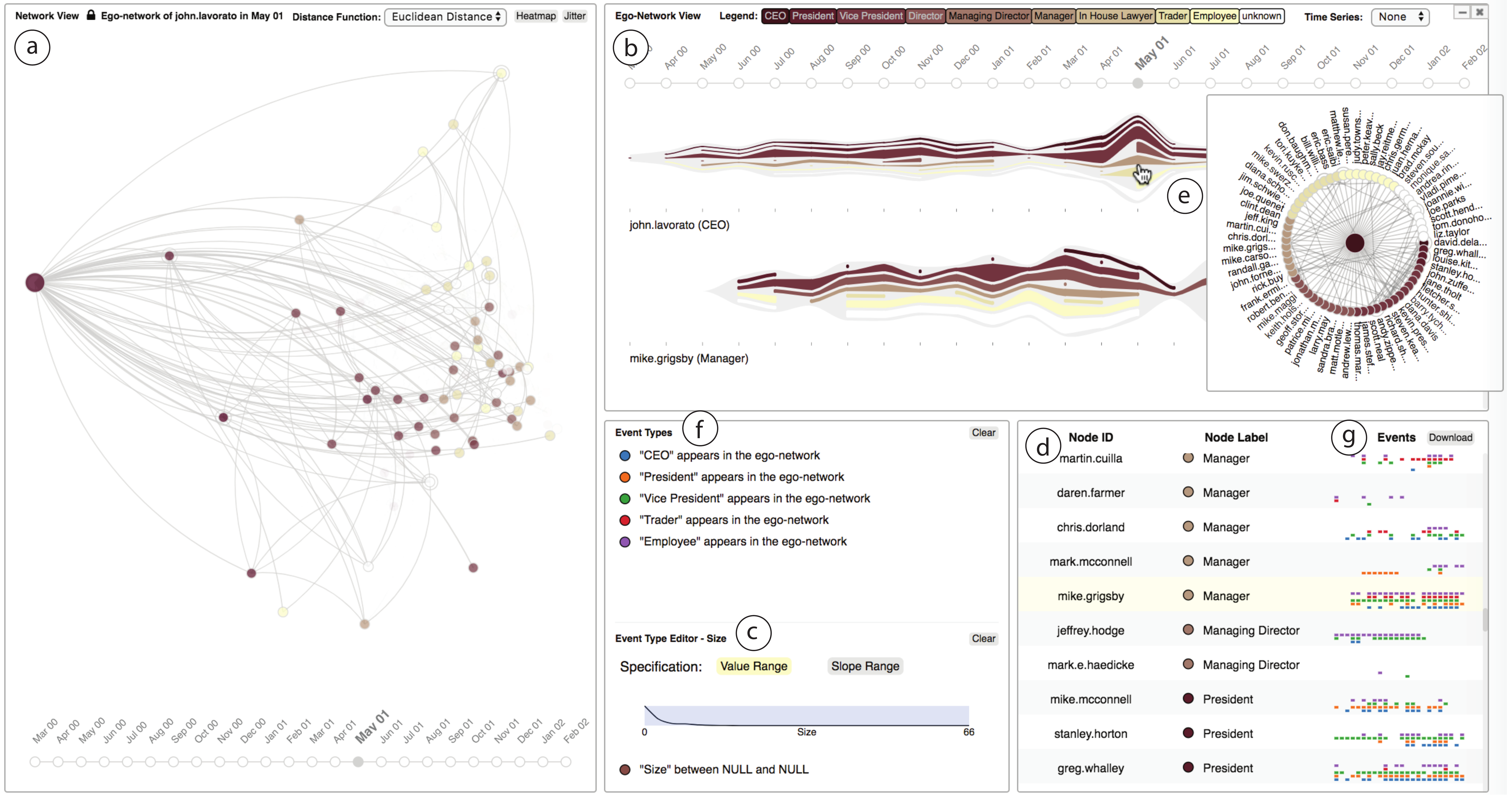}
  \centering
  \caption{Segue's user interface. (a) The network view shows a network snapshot in which nodes are positioned using MDS. (b) The ego-network view visualizes selected dynamic ego-networks as timeline visualizations. (c) The event type editor allows analysts to interactively convert dynamic ego-networks into event sequences for constructing different spatial layouts of dynamic ego-networks. (d) The table view is a list of all nodes and their labels. (e) The analyst hovers over the timeline to see a snapshot of an ego-network (here, ego-network of CEO John Lavorato in May 01). (f) A list of event types created by the analyst is shown above the event type editor. (g) Event sequences extracted from the ego-networks are visualized as pixel displays in the event column of the table view.}
  \label{teaser}
}

\begin{document}


\firstsection{Introduction}

\maketitle

Dynamic network analysis has been a topic of interest in many domains, including social networks \cite{socialnetworkexample}, scientific collaboration \cite{newman2001structure}, and healthcare \cite{healthcareexample}. Analyzing dynamic networks from the perspective of individuals provides insights into the behaviors of these individuals and the interactions among them over time. The analysis of individuals in a network context is referred to as egocentric network analysis or ego-network analysis. An ego-network consists of a focal node, the nodes within its one-step neighborhood and all the edges among these nodes \cite{sna}. A common task in ego-network analysis is to understand the overall evolution patterns of \textit{a collection of} ego-networks \cite{egoslider}. For instance, sociologists are interested in understanding whether people's social networks evolve incrementally or have a lot of sudden changes \cite{skypenetwork}. In healthcare policy, researchers attempt to determine if healthy people maintain larger social networks than unhealthy ones \cite{healthcareegonetwork}. These questions entail investigating the global picture of how a collection of ego-networks evolve to identify trends, outliers, and clusters of evolution patterns. They are often termed macroscopic-level questions \cite{egoslider}.

Answering macroscopic-level questions is challenging because analysts are required to gain an understanding of the evolution patterns of many dynamic ego-networks and draw connections among their evolution patterns. The complexity of exploring the evolution patterns of a single dynamic network was well-investigated in prior visualization research \cite{star, snapshot, dynamiccluster}: not only do analysts need to make sense of what happens in a static snapshot in a dynamic network, but they also need to gain insights into the ways a large number of static snapshots link together to form evolution patterns. A large number of dynamic ego-networks magnify the complexity by adding an additional dimension to the problem: each dynamic ego-network is itself a dynamic network and there are hundreds of dynamic ego-networks from which evolution patterns are to be understood.

The complexity involved in analyzing many ego-networks renders answering macroscopic-level questions difficult using existing dynamic network visualization techniques. Common techniques such as timeline and animation focus on visualizing the evolution patterns of a single dynamic network \cite{egoslider, egoline}. This is a mismatch with addressing macroscopic-level questions that requires getting a sense of the evolution patterns of many ego-networks. With these common techniques, analysts are forced to inspect the evolution patterns of a single ego-network using a timeline visualization or an animation, and look at the visualizations of each dynamic ego-network one after another. This process is tedious and cognitively demanding, especially when there are hundreds of ego-networks to review. Furthermore, analysts would find it difficult to mentally connect the evolution patterns of different dynamic ego-networks. An effective overview, on the other hand, allows analysts to see the evolution patterns of many dynamic ego-networks at once, thereby reducing analysts' effort to acquire the overall picture of how a collection of dynamic ego-networks evolve.

It has been shown that spatial layouts are well-suited for overviewing the patterns of a large number of data points by using the ``near=similar'' metaphor \cite{timeCurves, snapshot, SI}. In a spatial layout in which each dot represents a dynamic ego-network, clusters of dots indicate similar evolution patterns while outlying dots exhibit uncommon evolution patterns. A spatial layout can therefore enable analysts to see the evolution patterns of all dynamic ego-networks concurrently. In this study, we explore a technique that enables analysts to construct spatial layouts of dynamic ego-networks to answer macroscopic-level questions. This technique is developed with \textit{interpretability} and \textit{interactivity} in mind: analysts should be able to interpet why two points are close, and be able to interactively change the spatial layout to answer their evolving questions during analysis. To achieve this, we developed a data transformation pipeline that lets analysts convert dynamic ego-networks into event sequences to interactively create interpretable spatial layouts. Grounded in the pipeline, we developed Segue, a visual analysis system that empowers analysts to explore the evolution patterns of ego-networks by both seeing an overview of all dynamic ego-networks as well as examining the details of individual dynamic ego-networks. We demonstrate, through two usage scenarios, that analysts can interactively construct different spatial layouts to answer macroscopic-level questions. In particular, we make the following contributions:

\vspace{0.5mm}
\textbf{1.} A data transformation pipeline that enables analysts to \textit{interactively} create \textit{interpretable} spatial layouts of dynamic ego-networks to answer various macroscopic-level questions during analysis.

\vspace{0.5mm}
\textbf{2.} Segue, a visual analysis system that embodies the data transformation pipeline for exploring the overall evolution patterns of a collection of ego-networks.

\section{Related Work}

The design of Segue was informed by research in three areas: ego-network visualization, dynamic network visualization and techniques for creating spatial layouts for sensemaking.

\subsection{Ego-Network Visualization and Analysis}

Much work has been devoted to tailoring visualization techniques to ego-networks. Most of them focus on visualizing individual ego-networks rather than revealing the overall evolution patterns of a collection of them (e.g., \cite{egoline, 15d1, 15d2, quan}). One of the earliest works in this area is the tree-ring layout \cite{treeRing}, which uses tree rings to represent a dynamic ego-network. ManyNets \cite{manynets} visualizes an ego-network as a row in a table. Due to limited screen real estate, these techniques are limited to visualizing a dozen dynamic ego-networks at once. Zhao et al. \cite{egoline} proposed a subway map visualization to depict the evolution of a person's ego-network. Liu et al. \cite{egonetcloud} developed EgoNetCloud, which visualizes event-based egocentric dynamic network of an individual. MENA \cite{mena} and TMNVis \cite{tmnvis} investigated techniques to visualize multivariate ego-network evolution. While these techniques provide rich depictions of an individual ego-network, to get an overall picture of the evolution patterns of a large group of ego-networks, analysts are required to inspect ego-network one by one. Not only is this process tedious and cognitively demanding, it also fails to help analysts draw connections between the evolution patterns of different ego-networks. The most related work is egoSlider designed by Wu et al. \cite{egoslider}. They identified three levels of analysis questions in ego-network exploration: microscopic level, mesoscopic level and macroscopic level. Our work aims to address macroscopic-level questions. Macroscopic-level questions are often asked when analysts want to obtain a big picture of the evolution patterns of all the ego-networks (e.g., \cite{macro1, macro2, macro3, macro4, macroExample}). For instance, to understand the social interaction among Facebook users, Arnaboldi et al. \cite{macroExample} dived into the distributions of variables such as ego-network size and tie strength. To facilitate analysts to answer macroscopic-level questions, egoSlider uses a data overview that shows an MDS plot for each time step. Each dot in an MDS plot represents an ego-network and distance between dots encodes similarity. Wu et al. \cite{egoslider} commented that a limitation of this technique is it only allows analysts to see the similarity of ego-networks within one time step. Our work fills this gap by allowing analysts to interactively construct different spatial layouts in which a dot represents a dynamic ego-network. Closer dynamic ego-networks in the layout indicate that they share common evolution patterns. By providing an overview of the evolution patterns of ego-networks, our technique shields users from inspecting the evolution patterns of individual ego-networks one by one.

\subsection{Dynamic Network Visualization}

Another line of related work concerns general techniques for visualizing dynamic network. Beck et al. \cite{star} offered a comprehensive survey of different dynamic network visualization techniques. Two major approaches to analyze network evolution are animation and timeline. Animation-based technique uses animated transition of visual elements (e.g., nodes and edges in a node-link diagram) to reveal the time dimension. An obvious drawback is that it is cognitively demanding to keep track of the changes \cite{mentalMap}. Techniques such as staged animation (e.g., \cite{stage1, stage2}) are developed to help analysts preserve their mental maps and reduce the effort required to track changes between two consecutive time steps. However, it can still be difficult to focus on many items simultaneously and track changes over a long period of time \cite{track1, track2}. Answering macroscopic questions requires analysts to gain insights into how a large number of ego-networks evolve and how the evolutions of different ego-networks relate. Not being able to track changes of a large number of items makes these animation-based techniques unsuitable for revealing the global evolution patterns of a collection of ego-networks. Timeline-based approaches, on the other hand, use small multiples (e.g., \cite{smallMultiples}), vertical or horizontal timeline (e.g., \cite{timeline}) and circular layout (e.g., \cite{track2}) to represent the time dimension. However, as noted by Wu et al. these techniques mainly focus on tracking changes of the entire network rather than the characteristics of ego-networks \cite{egoslider}. Moreover, when there are many dynamic ego-networks, each needs to be shown using a timeline visualization. Analysts are required to inspect many of these timeline visualizations one by one to get a sense of the overall evolution patterns of many ego-networks. It is a time-consuming and cognitively-demanding process. We aim to go beyond timeline- and animation- based techniques to help analysts understand the global picture of a large number of ego-networks by allowing them to interactively construct an overview.

\subsection{Creating Spatial Layouts for Sensemaking}

Using the ``near=similar'' metaphor, spatial layouts have been known to be an effective technique for providing an overview of high-dimensional data \cite{SI}. An early work in interactive spatial layout construction is Dust \& Magnet \cite{DM}. It allows analysts to move magnets that represent attributes on a canvas. Data cases with a higher value are attracted to the magnets. This helps find patterns in a large number of multivariate data points in tabular data. Spatial layouts are also shown to be effective in analyzing text documents. Endert et al. \cite{SI} proposed a technique called semantic interaction that automatically infer parameters of statistical models from users' interactions with documents. The spatial layout of documents is automatically updated to reveal clusters of documents that are relevant to the analysis. Investigating how to adjust a spatial layout of objects to better reflect analysts' mental model and facilitate question answering has since become an active research area \cite{disfunction, interaxis, axisketcher, observationlevelinteraction}. Indeed, there has also been attempts to visualize temporal data as a spatial layout. For example, van den Elzen et al. \cite{snapshot} visualized a snapshot of a dynamic network as a point in a spatial layout and time curve \cite{timeCurves} represents a time point in temporal data as a point. We extend this line of research by contributing methods for interactively constructing interpretable spatial layouts in which each point represents a dynamic ego-network.

\section{An Interactive and Interpretable Technique for Constructing Spatial Layouts of Dynamic Ego-Networks}

In designing a technique for creating an overview of dynamic ego-networks, two desiderata are of importance: \textit{interpretability} \cite{interpret1, interpret2, interaxis} and \textit{interactivity} \cite{interaxis}. In our context, interpretability refers to the ability one can make inference from an overview. While an overview increases the scale of analysis by abstracting away low-level details, analysts may grapple with drawing useful inference from the overview due to information loss. For a spatial layout, understanding the fact that two near points share similar characteristics is not difficult. Interpreting in what way they are similar, however, can be a challenging task. Another consideration is interactivity, which refers to whether analysts can manipulate the overview to answer their diverse questions during data analysis. Spatial layouts are often generated by dimensionality reduction techniques (e.g., PCA \cite{pca}, MDS \cite{mds} and t-SNE \cite{tsne}) that are typically fully automated and do not allow human intervention \cite{interaxis}. Not being able to change the spatial layout, analysts are bound to answering their \textit{evolving} questions using a \textit{static} visualization, limiting the depth of exploration and curtailing thorough deliberation \cite{dynamics}.

Our goal is to develop an interactive and interpretable technique for constructing spatial layouts in which each point is a dynamic ego-network. While many techniques are designed to be interpretable and interactive (e.g., \cite{interaxis, axisketcher}), they do not to consider the time dimension of data points. To tackle \textit{interpretability}, we propose a data transformation pipeline that transforms dynamic ego-networks into event sequences. When analysts hover over a point in the spatial layout, a visualization of its event sequence is shown, thereby helping analysts to get a sense of, for example, the unique evolution patterns of an ego-network that make it an outlying point. To address \textit{interactivity}, our technique allows analysts to adjust pairwise distances between dynamic ego-networks by selecting the characteristics of dynamic ego-networks relevant to the analysis. This changes the distance matrix and hence produces various spatial layouts to reveal different global evolution patterns. In this section, we first describe the data model supported by Segue. We then illustrate the data transformation pipeline in greater detail.

\subsection{Data Model}

Segue supports analysis of a collection of dynamic ego-networks. A dynamic ego-network with a focal node $i$ can be represented as $\Gamma_{i}=\{G_{i}^{1},...,G_{i}^{N}\}$, where $N$ is the total number of time steps. An ego-network with a focal node $i$ at time step $t$ is modeled as an undirected graph $G_{i}^{t}$ with a set of nodes $V_{i}^{t}$ and a set of edges $E_{i}^{t}$. Each node in $V_{i}^{t}$ has an attribute. Throughout the remainder of our study, we use the widely-studied Enron email network\cite{enron} as a running example. There are 142 employees in our Enron dataset. Each individual has a dynamic ego-network. An ego-network snapshot depicts the email communication of an employee with other employees in a given month. In a snapshot of an employee's ego-network, another employee is present if there are email exchanges between them. An edge in the snapshot represents an email exchange. Each node has an attribute, which captures the highest position attained by that employee (e.g., CEO, President, and Employee). The dataset spans the 24-month between Mar 2000 and Feb 2002 during which Enron went bankrupt. Hence, there are 142 dynamic ego-networks (corresponding to the number of individuals), each having 24 snapshots (corresponding to the timespan).

\begin{figure}[h!]
  \centering
  \vspace{-1mm}
  \includegraphics[width=\linewidth]{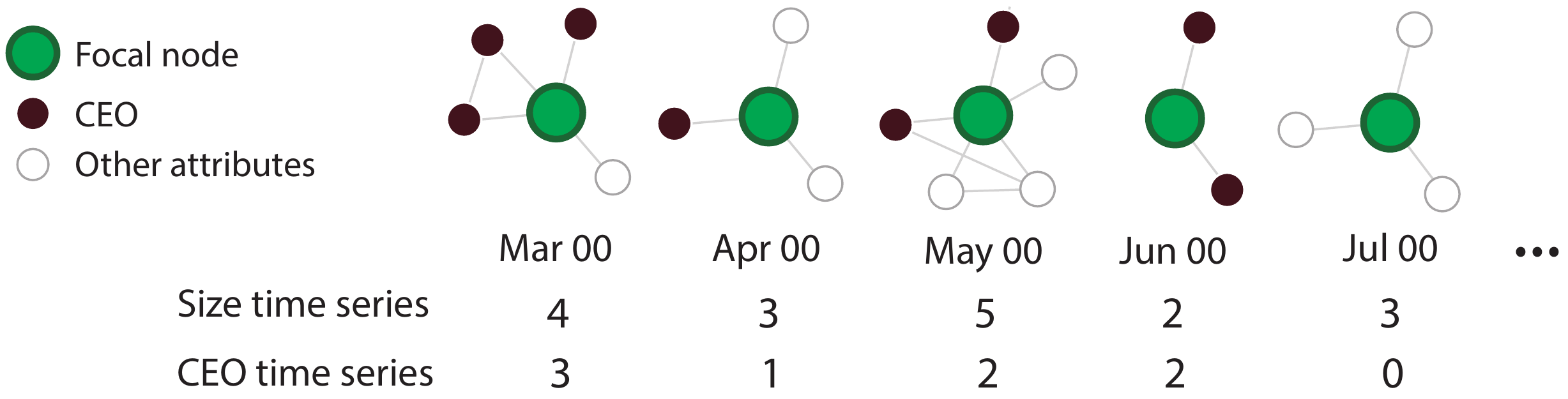}
  \vspace{-6mm}
  \caption{A dynamic ego-network in the Enron email network. It is represented as multiple time series (e.g., size and CEO) in Segue.}
  \label{egoNet}
\end{figure}

To better summarize the evolution patterns of ego-networks, each dynamic ego-network consists of a set of time series. Each time series in the set is a series of numerical values derived either from node attributes or some structural properties of a dynamic ego-network. A pictorial description of how Segue represents one dynamic ego-network is shown in Figure ~\ref{egoNet}. Examples of time series for a dynamic ego-network in the Enron dataset are a size time series that describes the number of nodes to which the focal node is linked over time and a CEO time series that describes the number of CEOs to which a focal node is connected over time. Each time series also has a type $T_{i}\in\{ T_{1},...,T_{M} \}$, where $M$ is the total number of time series types. For instance, CEO is the type of the CEO time series of each dynamic ego-network. The following is a full list of time series types in the Enron dataset. Without loss of generality, time series types that describe other properties of dynamic ego-networks (e.g., tie strength) can be added to the list.

\begin{figure}[h!]
  \centering
  \vspace{-1mm}
  \includegraphics[width=0.76\linewidth]{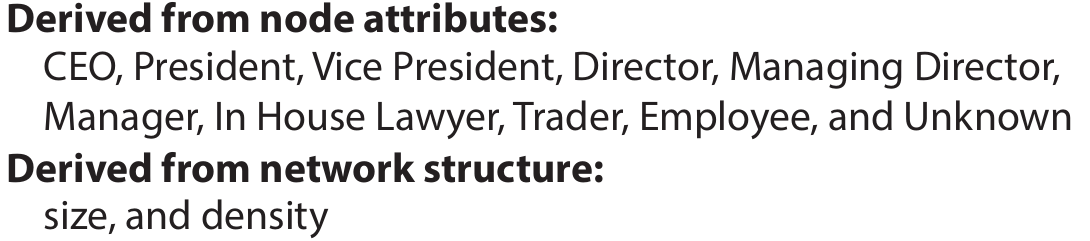}
  \vspace{-3mm}
\end{figure}

\subsection{Segue's Data Transformation Pipeline}

\begin{figure}[h!]
  \centering
  \vspace{-2mm}
  \includegraphics[width=\linewidth]{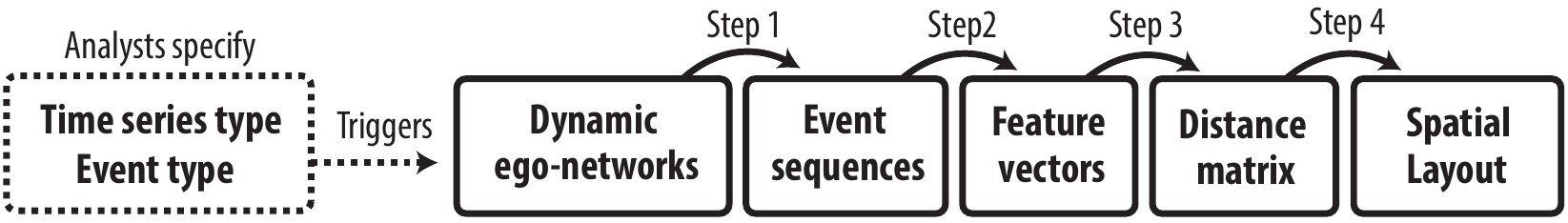}
  \vspace{-7mm}
  \caption{An overview of Segue's data transformation pipeline.}
  \vspace{-2mm}
  \label{pipeline}
\end{figure}

Segue's data transformation pipeline consists of four major steps (Fig.~\ref{pipeline}). Analysts first select a time series type and define an event type. This triggers the \textbf{\textit{transformation of dynamic ego-networks into event sequences}}. The pipeline then \textbf{\textit{converts each event sequence into a feature vector}}. \textbf{\textit{The distance between each pair of dynamic ego-networks is computed}} as the distance between their feature vectors. Having computed a distance matrix, \textbf{\textit{the spatial layout of dynamic ego-networks is updated using MDS}}.

\subsubsection*{Step 1: Transforming Dynamic Ego-Networks \\ into Event Sequences}

\begin{figure}[h!]
  \centering
  \vspace{-4mm}
  \includegraphics[width=\linewidth]{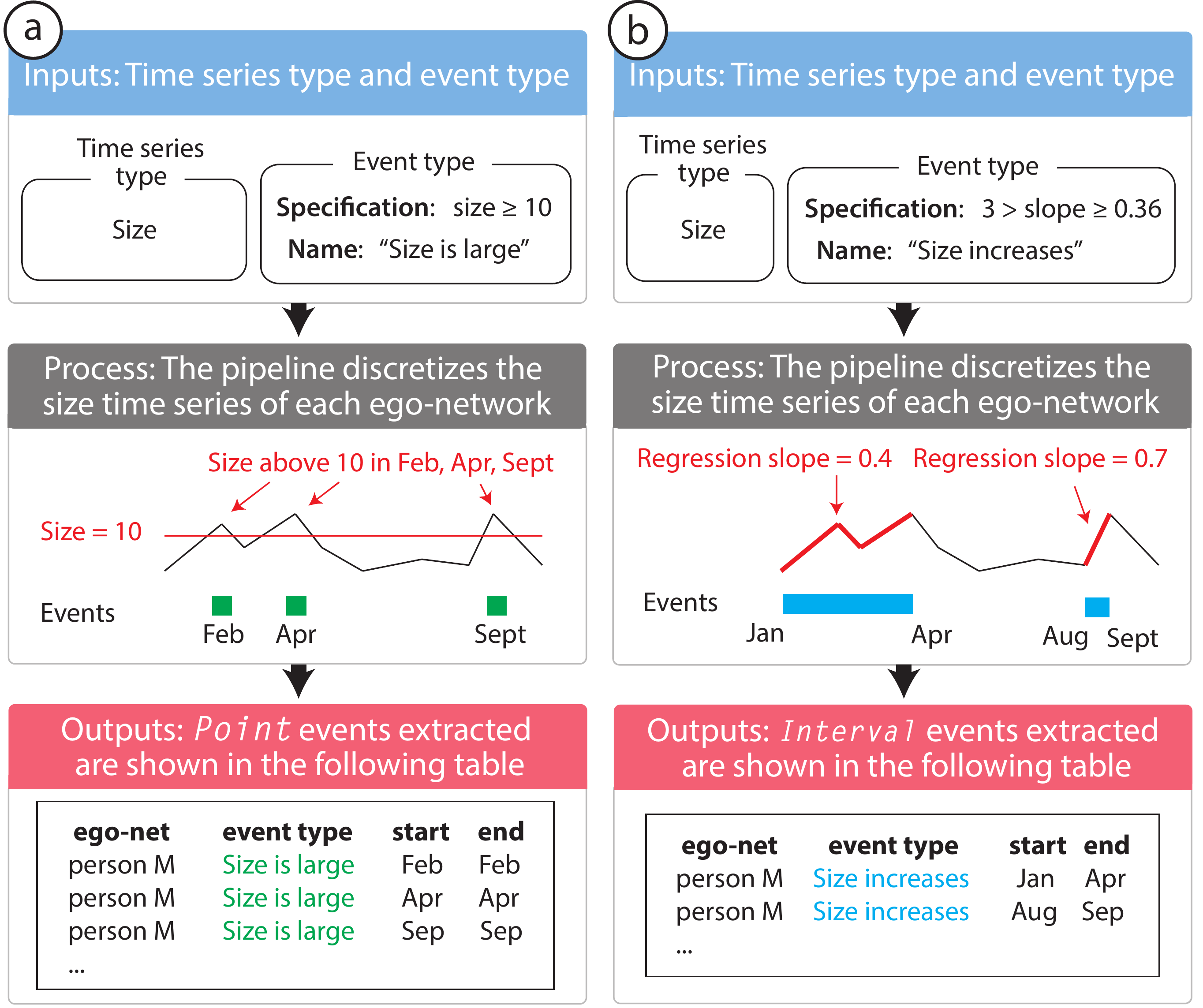}
  \vspace{-7mm}
  \caption{The inputs, process and outputs for extracting events from dynamic ego-networks. (a) Point events are created when the event type's specification is a range of values (e.g., size $\geqslant$ 10). (b) Interval events are created when the specification is a range of slopes (e.g., 3 $>$ slope $\geqslant$ 0.36).}
  \label{process}
\end{figure}

The input, process and output of how dynamic ego-networks are transformed into event sequences are shown in Figure~\ref{process}. The data transformation pipeline requires two inputs from users: a time series type and an event type. An event type consists of two attributes: a specification and a name. The specification of an event type is either a range of values or a range of slopes of time series, and the name is defined by users. The pipeline extracts events by discretizing the specified type of time series of each dynamic ego-network based on the specification of the event type. Converting dynamic ego-networks into event sequences offers the possibility for enhancing the \textit{interpretability} of the spatial layout: as analysts hover over a point in the spatial layout, the event sequences extracted by analysts can be visualized to help analysts interpret, for example, why two points are near.

The pipeline generates point events if the specification of the event type is a range of values and generates interval events if the specification is a range of slopes. Point events occur at a point in time and interval events span over a period of time \cite{eventflow}. To illustrate, suppose an analyst has selected size as the time series type. When she sets the specification of an event type to be ``size $\geqslant$ 10'' and names the event type as ``size is large'' (Fig.~\ref{process}a), the algorithm creates a point event for each time point in an ego-network's size time series where the size value is greater than 10. All these point events have the type ``size is large''. On the other hand, if she specifies an event type named ``size increases'' with a specification ``3 $>$ slope $\geqslant$ 0.36'' (Fig.~\ref{process}b), the algorithm generates interval events for time intervals in a size time series within which the regression slopes are between 0.36 and 3. The interval events extracted have the type ``size increases''. A constraint in interval event extraction is that interval events should not overlap (The pseudocode of the interval event extraction algorithm is outlined in the supplementary document). 

More formally, the pipeline transforms the $i^{th}$ time series $T_{ij}$ of a dynamic ego-network $j$ into a set of events $\{e^{A}_{ijsd}\}$ of type $E_{A}$. Each event $e^{A}_{ijsd}$ of type $E_{A}$ is extracted from the time series $T_{ij}$ between time $t=s$ and $t=d$. Both the time series type $T_{i}$ and event type $E_{A}$ are specified by analysts.

\subsubsection*{Step 2: Converting Event Sequences into Feature Vectors}

\begin{figure}[h!]
  \centering
  \vspace{-2mm}
  \includegraphics[width=0.9\linewidth]{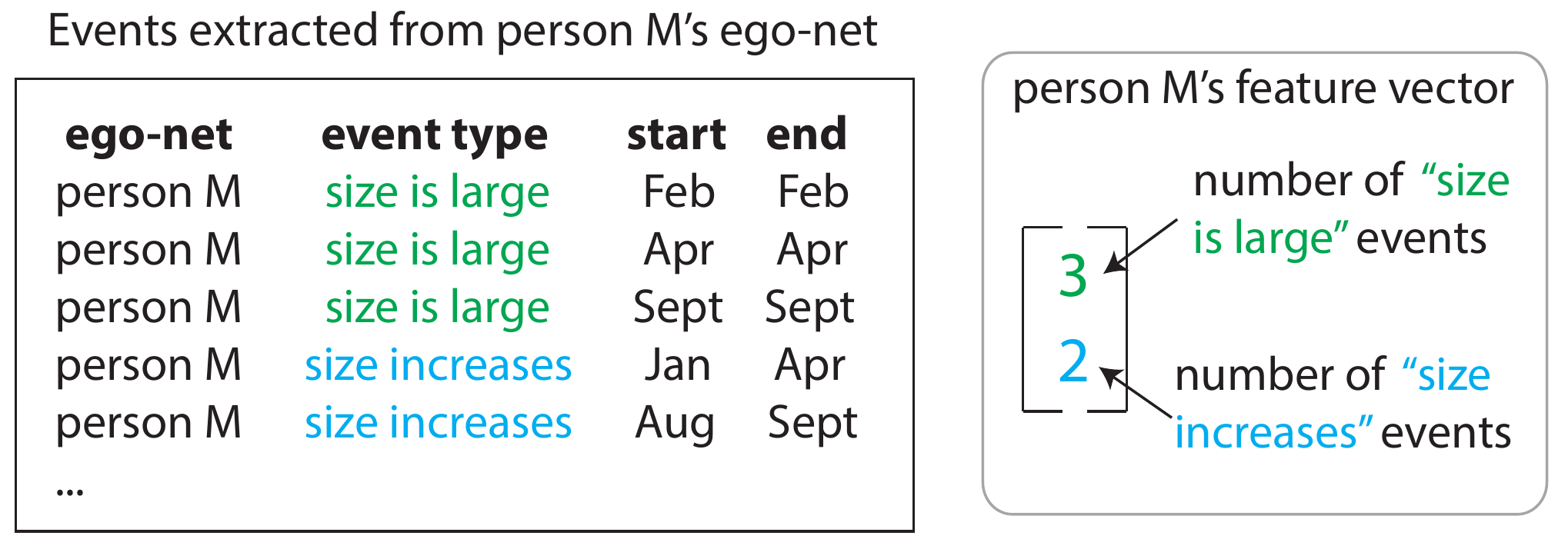}
  \vspace{-3mm}
  \caption{Converting the event sequence extracted from a dynamic ego-network into a feature vector.}
  \vspace{-2mm}
  \label{featureVector}
\end{figure}

After transforming dynamic ego-networks into event sequences, the data transformation pipeline converts the event sequences into feature vectors. Figure~\ref{featureVector} shows an example of the conversion. In Figure~\ref{featureVector}, the first and second features in person M's feature vector correspond to the event type ``size is large'' and ``size increases'' respectively. The first feature has a value of three as events of type ``size is large'' appears three times in person M's event sequence. By converting dynamic ego-networks into feature vectors, this step helps compute the distances between dynamic ego-networks for constructing a spatial layout.

More formally, suppose an analyst has defined a set of $n$ event types $E=\{E_{1},...,E_{n}\}$. The extracted event sequence $S_{j}=(e_{1},...,e_{m})$ of a dynamic ego-network $j$ is an ordered list of events, where $e_{i} \in E$. The pipeline converts the event sequence $S_{j}$ of a dynamic ego-network $j$ into an $n$-dimensional feature vector $\vec{V_{j}}=<v_{1},...,v_{n}>$. The $i$-th element in $\vec{V_{j}}$ is the count of events of type $E_i$ in $S_{j}$.

\subsubsection*{Step 3: Computing Pairwise Distances between Ego-Networks}

After converting each event sequence into a feature vector, the data transformation pipeline computes the distance between each pair of dynamic ego-networks as the Euclidean distance between the corresponding pair of feature vectors:

\vspace{-1mm}
\[ d(\Gamma_{i},\Gamma_{j})=||\vec{V_{i}}-\vec{V_{j}}|| \]
\vspace{-5mm}

\noindent We compute pairwise distances by transforming event sequences into feature vectors because this method was proved effective by other systems such as EventAction \cite{eventaction}. With the pairwise distances between the dynamic ego-networks, a distance matrix in which each row/column represents a dynamic ego-network can be computed.

We note that Euclidean distance should only be considered a baseline distance function for demonstrating the utility of the data transformation pipeline. A clear drawback of Euclidean distance is that it does not consider the temporal aspect of event sequences. For example, two sequences {\small \textbf{\texttt{[A, B, A, B, A]}}} and {\small \textbf{\texttt{[A, A, A, B, B]}}}, where {\small \textbf{\texttt{A}}} and {\small \textbf{\texttt{B}}} are events, have the same feature vector and hence a Euclidean distance of zero. As noted by Du et al. \cite{peerfinder}, distance between event sequences can be subjective and depends on application domains. To deal with more advanced applications that require distance function beyond Euclidean distance, step 2 and 3 can be replaced by other methods of computing distances between two event sequences. An example is edit distance \cite{edit2}, the minimum number of operations required to transform one sequence into another sequence. Edit distance considers the temporal aspect of event sequences and is often used as a distance function for event sequences \cite{edit}. We will review other possible enhancements in Section~\ref{limitation}.

\subsubsection*{Step 4: Generating a Spatial Layout of Dynamic Ego-Networks}

With a distance matrix, we use classical MDS to project dynamic ego-networks onto a spatial layout. We choose classical MDS because of its fast convergence \cite{mds}. Fast convergence makes \textit{interactivity} possible: every time when analysts adjust the distance matrix by specifying a time series type and an event type, Segue renders the spatial layout immediately. Step 4 can be replaced by other dimensionality reduction algorithms that take as input a distance matrix. These algorithms include force-directed MDS and t-SNE. We have performed informal testings on these algorithms and observed that they were slower and therefore less scalable for rendering a spatial layout of many dynamic ego-networks.

The spatial layout generated reveals the evolution patterns of the dynamic ego-networks: the more similar the evolution patterns are, the closer the dynamic ego-networks are in the spatial layout. This is because two dynamic ego-networks tend to have small distance if most event types appear similar number of times in their event sequences. Each event type defined by analysts encodes some information about a particular evolution pattern. If two ego-networks share similar evolution patterns, they will have similar number of events of the same type, thereby pulling them closer together in the spatial layout. For instance, to move apart the dynamic ego-networks that are fluctuating in size from those that are fairly stable in size in the spatial layout, an analyst can create two event types with the following names and specifications:

\begin{figure}[h!]
  \centering
   \vspace{-2.5mm}
  \includegraphics[width=0.9\linewidth]{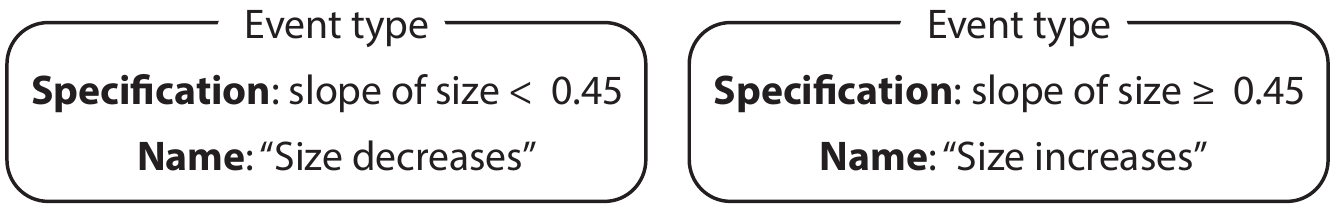}
   \vspace{-2.5mm}
\end{figure}

\begin{figure*}[t]
  \centering
  \includegraphics[width=\linewidth]{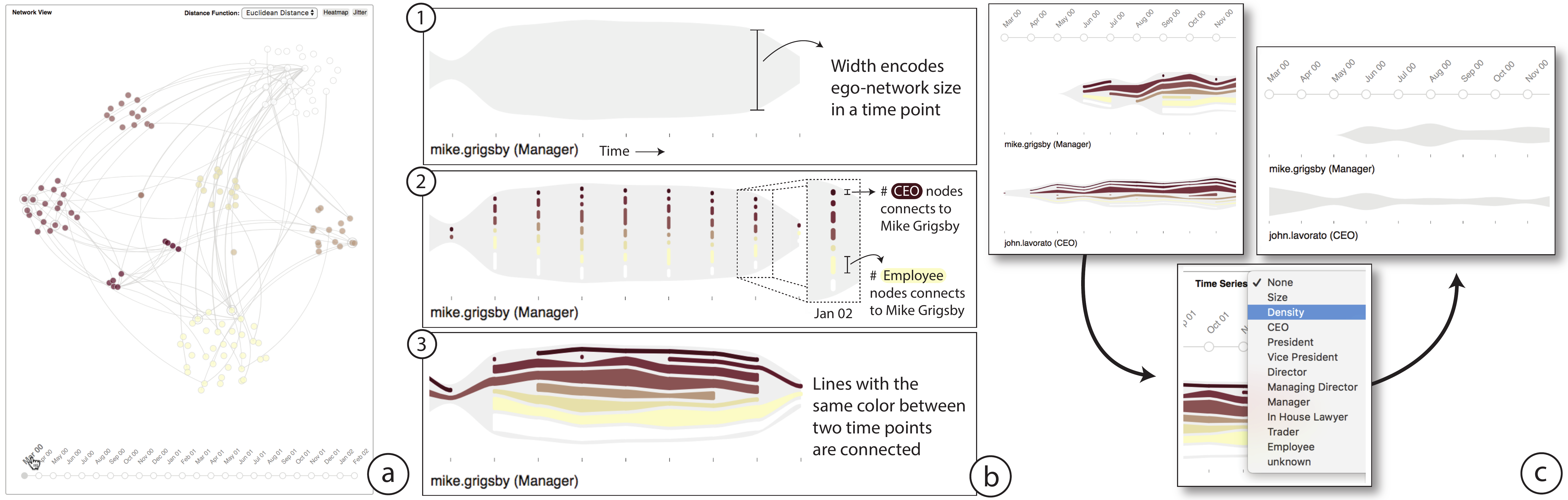}
  \vspace{-7mm}
  \caption{Visualization and interaction designs in the network view and the ego-network view. (a) The analyst hovers over ``Mar 00'' to see all the edges in that month. (b) The three-step construction of the timeline visualization. (c) The analyst selects density from the drop-down menu to convert the timeline visualizations in the ego-network view into area charts which represents the density time series.}
  \vspace{-4mm}
  \label{vis}
\end{figure*}

\noindent Both event types encode the analyst's definition of fluctuating in size. If two dynamic ego-networks are highly fluctuating in size, they will have more events of both types and hence will be closer in the spatial layout. On the contrary, if one dynamic ego-network is fluctuating in size but another one is stable, the former will have many more events of both types than the latter, pushing them farther away in the spatial layout. Therefore, closeness in the resulting spatialization indicates how similar the evolution patterns of dynamic ego-networks are.

\section{Segue's User Interface}

We followed the principles of iterative design \cite{norman} in designing Segue (images of previous prototypes are provided in the supplementary materials). Segue was designed to enable analysts to explore evolution patterns of ego-networks by \textit{interactively} constructing various \textit{interpretable} spatial layouts using the aforementioned data transformation pipeline. Aside from creating a system that does what the pipeline can do, we also considered other requirements in the process of iterative design. To gather design requirements, we conducted a formative evaluation of the initial prototype with two experts. The expert reviews were designed based on the guidelines in \cite{expertreview}. The first expert is a graph expert with $>$10 years of experience in graph research and recent experience in event sequence research. The second expert is an event sequence expert with $\sim$3 years of event sequence research experience and some experience in dynamic network research. We recruited experts who understand event sequence data because event sequences are one of the core components in our system. Each interview lasted for approximately an hour. We first provided the project background and a tutorial of how the prototype worked. We then described illustrative use cases based on the Enron dataset to help our experts determine how well the tool met users' needs. After that, we let them explore the tool freely. Lastly, we collected some initial feedback from them. Post-interview, we provided the experts remote access to the prototype as well as follow-on questions for additional feedback. We report the considerations for designing the final version of Segue below:

\vspace{1mm}
\noindent \textbf{C1. Enable thorough exploration of dynamic ego-networks.} Besides creating different spatial layouts for getting a high-level overview of ego-network evolution, analysts may want to inspect the low-level details of each dynamic ego-network. Segue should visualize low-level details, such as the time series by which a dynamic ego-network is represented and the ego-network snapshots at different time points.

\vspace{1mm}
\noindent \textbf{C2. Provide mechanisms for verifying the transformation.} When analysts define an event type, the data transformation pipeline discretizes a time series of each dynamic ego-network into an event sequence based on the event type's specification. Analysts may want to verify whether the transformation is done correctly. Segue should allow analysts to verify the transformation from time series into event sequences.

\vspace{1mm}
\noindent \textbf{C3. Facilitate interpretation of the spatial layout.} The spatial layout in which dots represent dynamic ego-networks provides information about how similar the evolution patterns of the ego-network are but not in what way the evolution patterns are similar. For example, close dots indicate that the ego-networks share common evolution patterns but what the common patterns are cannot be identified directly from the spatial positions. Segue should offer some mechanisms for interpreting the spatial positions of dynamic ego-networks.

\vspace{1mm}
\noindent \textbf{C4. Maintain consistency of timelines in different views.} Consistency is an important consideration for designing visualization systems \cite{consistent}. In the initial prototype, the timelines in different views were oriented in different directions. The experts commented that as the timelines shared the same time scale, they should be consistently oriented in the same direction. 

\subsection{Basic Visualization and Interaction Designs}

Besides allowing analysts to interactively create spatial layouts, Segue contains a network view and an ego-network view to facilitate thorough exploration of evolution patterns of ego-networks (C1).

\subsubsection{Network View}

The network view serves dual functions. First, it is a spatial layout in which each dot represents a dynamic ego-network. At the beginning, when analysts have not created any event types, Segue puts the dots into clusters based on the attributes of the focal node (Fig.~\ref{vis}a). As analysts add event types to convert the dynamic ego-networks into event sequences, Segue computes the distances between dynamic ego-networks based on their event sequences and lays out the dots using MDS.

Second, the network view is also a viewer of the bigger dynamic network that subsumes all the dynamic ego-networks. Hovering over a time point on the bottom timeline shows the links among the nodes at that time point (Fig.~\ref{vis}a). The nodes are colored by a node attribute. For the Enron dataset, each node represents an employee and its color encodes the employee's rank in Enron (the higher the rank, the darker the color). Links represents email communications among employees.

To construct spatial layouts using different distance functions, users can also choose between Euclidean distance, which does not consider the temporal aspect of event sequences and edit distance, which considers the temporal aspect of event sequences from the distance function menu (Fig.~\ref{teaser}a top right).

\subsubsection{Ego-Network View}

Being a high-level overview of the overall evolution patterns of ego-networks, the spatial layout in the network view only provides information about ego-network closeness and abstracts away other information. To tackle this information loss, we designed the ego-network view to provide the details of an individual ego-network's evolution (C1). Analysts can select an ego-network by clicking on a node in the network view or a row in the table view. The dynamic ego-network with the selected node as the focal node is then shown as a timeline visualization in the ego-network view (Fig.~\ref{teaser}b).

Segue constructs this timeline visualization in three steps (Fig.~\ref{vis}b). First, a gray area chart is created as the background. Its width encodes the ego-network size (i.e. number of nodes in the ego-network excluding the focal node) at each time point. Next, at each time point, some lines are drawn to indicate the number of nodes with different node attributes that are connected to the focal node. Finally, the lines that have the same color between two time points are connected. Linking the lines helps analysts identify whether the focal node connects to nodes with a particular attribute continuously.

Segue normalizes each timeline visualization by the maximum size of the ego-network. As the maximum size of different ego-networks can be drastically different, without normalization, small ego-networks will have a very narrow gray area chart. There will not be enough space for squeezing in the lines and links.

Segue represents each dynamic ego-network as multiple time series. Analysts can visualize a time series of an ego-network by selecting a time series type (e.g., CEO) from the time series drop-down menu (C1). The timeline visualizations in the ego-network view are then converted to area charts (Fig.~\ref{vis}c). The width of an area chart shows how the time series fluctuates over time. We use area chart rather than the commonly-used line chart to be consistent with the timeline visualization, which also used an area chart as the background.

As analysts hover over a timeline visualization, a node-link diagram depicting the ego-network snapshot at the selected time point is shown in a small window close to the mouse cursor (Fig.~\ref{teaser}e). The ego-network at the selected time point is also highlighted in the network view (Fig.~\ref{teaser}a). The largest node in both the small window and the network view is the focal node. By clicking on a timeline visualization, analysts store the selected ego-network in the network view. This keeps the ego-network highlighted in the network view even when the mouse cursor leaves the timeline visualization. Visualizing both dynamic ego-networks as timeline visualizations, and the snapshots in dynamic ego-networks as node-link diagrams allows thorough exploration of evolution patterns (C1).

\subsection{Interactively Constructing Spatial Layouts}

To walk through the process of creating a spatial layout using Segue, we describe how an analyst can create a spatial layout in which (S1) the dynamic ego-networks that maintain a similar size are closer and (S2) the dynamic ego-networks has a similar degree of fluctuation in size are closer. To achieve this, the analyst is going to create four event types: \raisebox{-0.3\height}{\includegraphics[scale = 0.3]{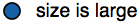}}, \raisebox{-0.3\height}{\includegraphics[scale = 0.3]{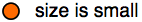}}, \raisebox{-0.3\height}{\includegraphics[scale = 0.3]{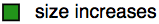}} and \raisebox{-0.2\height}{\includegraphics[scale = 0.3]{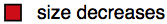}}.

\subsubsection{Specifying Event Types with a Range of Values}

As the analyst wants to separate the dynamic ego-networks that maintain a large size from those that maintain a small size (S1), she creates \raisebox{-0.2\height}{\includegraphics[scale = 0.3]{5_2_sizeislarge}} and \raisebox{-0.15\height}{\includegraphics[scale = 0.3]{5_2_sizeissmall}}. The specifications of both event types are ranges of values.

\begin{figure}[h!]
  \centering
  \vspace{-2mm}
  \includegraphics[width=\linewidth]{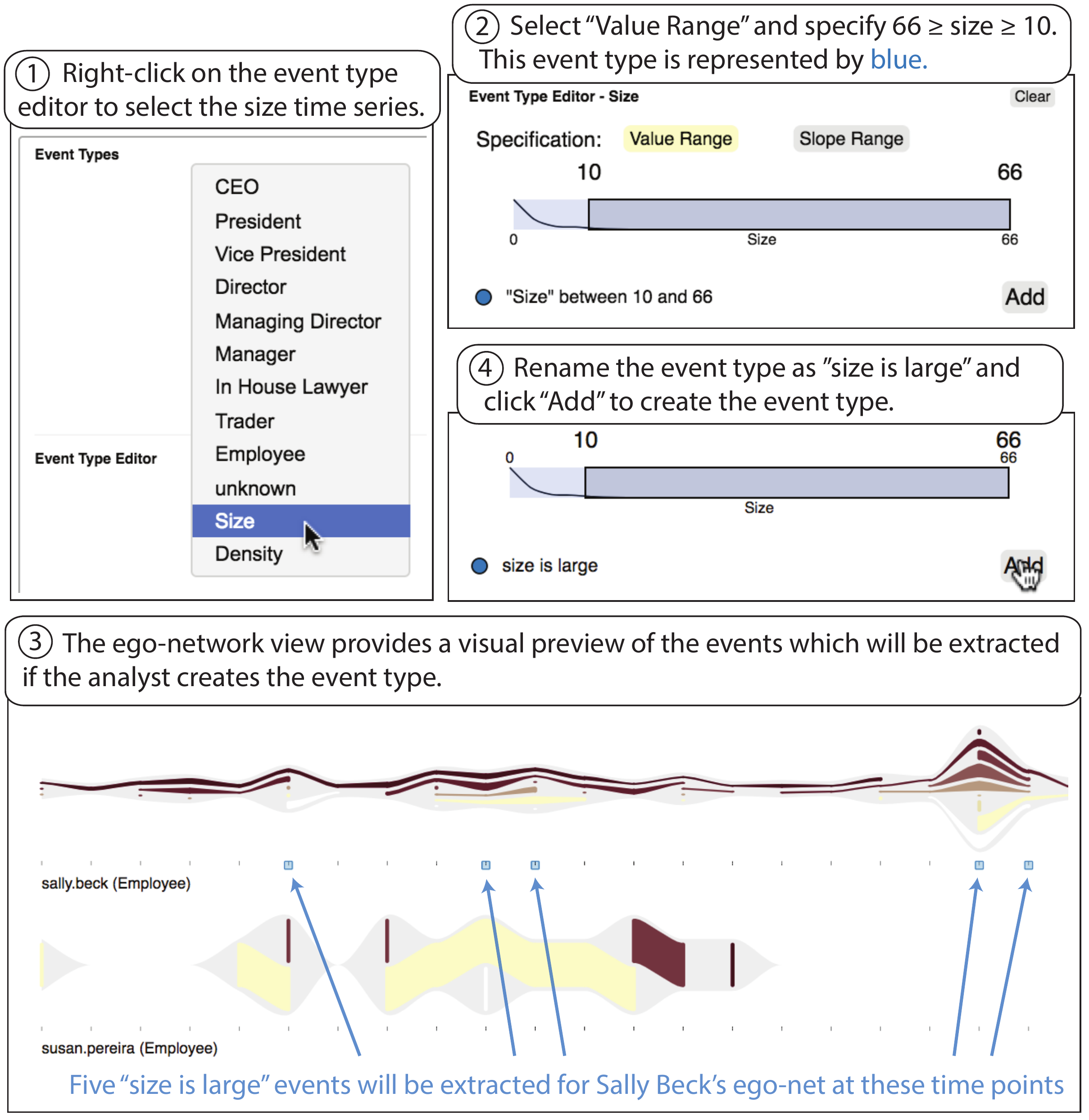}
  \vspace{-6.5mm}
  \caption{Specifying an event type with a range of values.}
  \vspace{-1mm}
  \label{eventType}
\end{figure}

To create \raisebox{-0.3\height}{\includegraphics[scale = 0.3]{5_2_sizeislarge}}, she right-clicks on the event type editor and selects the size time series from the context menu (Fig.~\ref{eventType}\raisebox{-0.25\height}{\includegraphics[scale = 0.25]{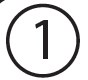}}). She then selects \raisebox{-0.3\height}{\includegraphics[scale = 0.6]{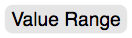}} from \raisebox{-0.3\height}{\includegraphics[scale = 0.65]{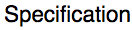}} (Fig.~\ref{eventType}\raisebox{-0.25\height}{\includegraphics[scale = 0.25]{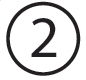}}). A data visualization slider \cite{visslider} is displayed for selecting a range of values. From the slider, the analyst notices that the minimum ego-network size is zero (a focal node does not connect to any nodes) and the maximum size is 66. The line chart in the data visualization slider shows the number of ego-network snapshots with a particular size. The analyst observes that most of the ego-network snapshots have a very small size. She then specifies a range, 66 $>$ size $\geqslant$ 10. As she specifies the range, Segue provides a visual preview of the events that will be created for the dynamic ego-networks in the ego-network view. The preview helps analysts to adjust the range before committing to creating the event type (C2). The preview in Figure~\ref{eventType}\raisebox{-0.25\height}{\includegraphics[scale = 0.25]{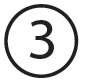}} tells the analyst that if she defines an event type by specifying 66 $>$ size $\geqslant$ 10, five point events will be created for Sally Beck's dynamic ego-network and no event will be created for Susan Pereira's dynamic ego-network. The analyst names the event type as ``size is large'' and click on \raisebox{-0.3\height}{\includegraphics[scale = 0.25]{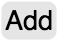}} to create \raisebox{-0.2\height}{\includegraphics[scale = 0.3]{5_2_sizeislarge}}, which is represented by blue in Segue's interface. Segue then extracts the point events according to the range specified. All the events extracted have the type \raisebox{-0.3\height}{\includegraphics[scale = 0.3]{5_2_sizeislarge}}. The analyst follows a similar procedure to create \raisebox{-0.3\height}{\includegraphics[scale = 0.3]{5_2_sizeissmall}} by specifying the range, size $<$ 3.

\begin{figure}[h!]
  \centering
  \vspace{-1mm}
  \includegraphics[width=\linewidth]{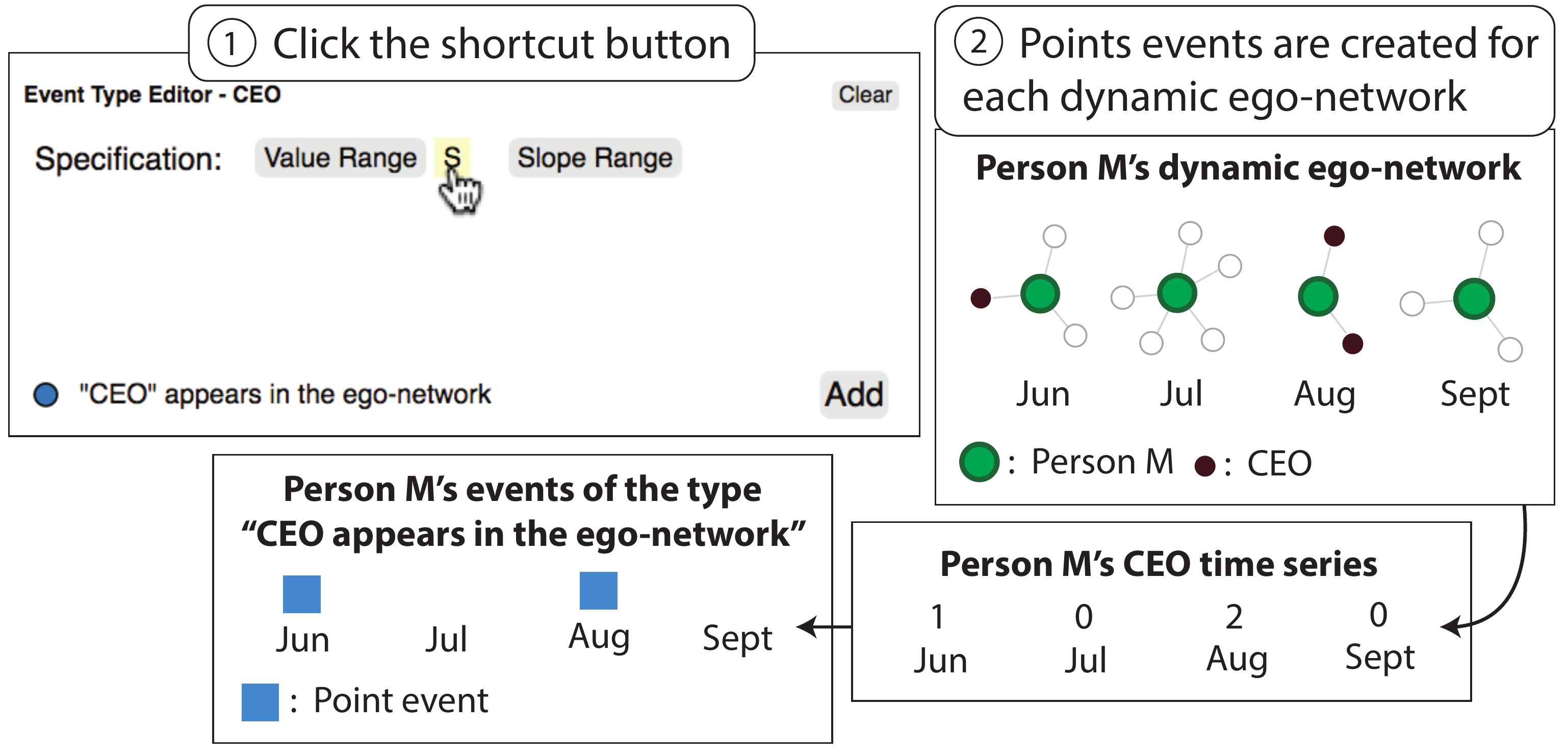}
  \vspace{-7mm}
  \caption{Shortcut to specifying an event type with a range of values.}
  \vspace{-2mm}
  \label{shortcut}
\end{figure}

For time series derived from node attributes (e.g., CEO and Employee), Segue supports a shortcut to specifying event types with a value range. Analysts use the shortcut by first selecting a time series (e.g., CEO) from the context menu and click on \raisebox{-0.3\height}{\includegraphics[scale = 0.6]{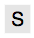}} from \raisebox{-0.3\height}{\includegraphics[scale = 0.65]{5_2_1_specification}} (Fig.~\ref{shortcut}\raisebox{-0.25\height}{\includegraphics[scale = 0.25]{one}}). Analysts then click on \raisebox{-0.3\height}{\includegraphics[scale = 0.25]{5_2_1_add}} to add the event type. If the CEO time series is selected, for example, Segue creates a point event for a time step as long as there is a CEO node (number of CEO nodes $\geqslant$ 1) in the ego-network at that time step (Fig.~\ref{shortcut}\raisebox{-0.25\height}{\includegraphics[scale = 0.25]{two}}).

\subsubsection{Specifying Event Types with a Range of Slopes}

In order to separate the dynamic ego-networks that has a fluctuating size from those that are stable in size (S2), the analyst creates \raisebox{-0.2\height}{\includegraphics[scale = 0.3]{5_2_sizeincreases}} and \raisebox{-0.2\height}{\includegraphics[scale = 0.3]{5_2_sizedecreases}}. The specifications of both event types are ranges of slopes.

\begin{figure}[h!]
  \centering
  \vspace{-1mm}
  \includegraphics[width=0.9\linewidth]{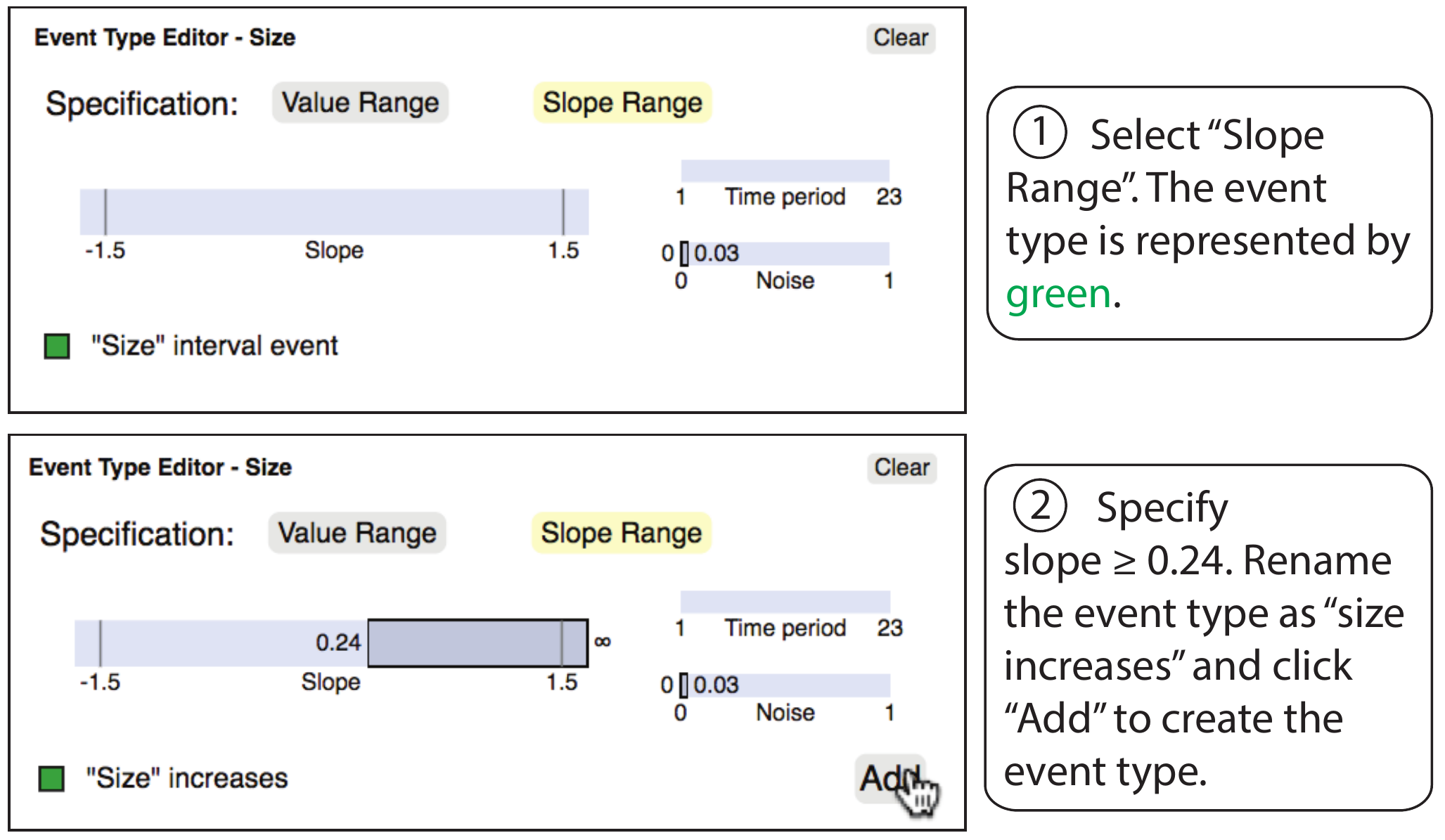}
  \vspace{-3mm}
  \caption{Specifying an event type with a range of slopes.}
  \vspace{-3mm}
  \label{slope}
\end{figure}

To create \raisebox{-0.3\height}{\includegraphics[scale = 0.3]{5_2_sizeincreases}}, she selects \raisebox{-0.3\height}{\includegraphics[scale = 0.6]{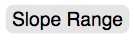}} from \raisebox{-0.3\height}{\includegraphics[scale = 0.65]{5_2_1_specification}} (Fig.~\ref{slope}\raisebox{-0.25\height}{\includegraphics[scale = 0.25]{one}}) and specifies the range of slopes to be slope of size $\geqslant$ 0.5 using the slope slider. The ego-network view provides a preview of the events that will be extracted if the analyst commits to creating the event type. After inspecting the preview, she wants to set a larger range of slope. She adjusts the range to slope of size $\geqslant$ 0.24. She then clicks on \raisebox{-0.3\height}{\includegraphics[scale = 0.25]{5_2_1_add}} to create \raisebox{-0.2\height}{\includegraphics[scale = 0.3]{5_2_sizeincreases}}, which is represented by green in the interface (Fig.~\ref{slope}\raisebox{-0.3\height}{\includegraphics[scale = 0.25]{two}}). To create \raisebox{-0.2\height}{\includegraphics[scale = 0.3]{5_2_sizedecreases}}, she follows similar steps and specifies slope of size $<$ -0.24.

\subsubsection{Interface Updates after Defining Event Types}

The list of all event types created is displayed above the event type editor (Fig.~\ref{teaser}f). A circle next to an event type name (e.g., \raisebox{-0.3\height}{\includegraphics[scale = 0.3]{5_2_sizeislarge}}) indicates that all the events which belong to this type are point events. A square next to an event type name (e.g., \raisebox{-0.3\height}{\includegraphics[scale = 0.3]{5_2_sizeincreases}}) means that all the events which belong to this type are interval events. Analysts can double-click on an event type in the event type editor to remove it. As analysts add event types to or remove event types from the list, the event sequences of the dynamic ego-networks are changed.

\begin{figure}[h!]
  \centering
  \vspace{-1mm}
  \includegraphics[width=\linewidth]{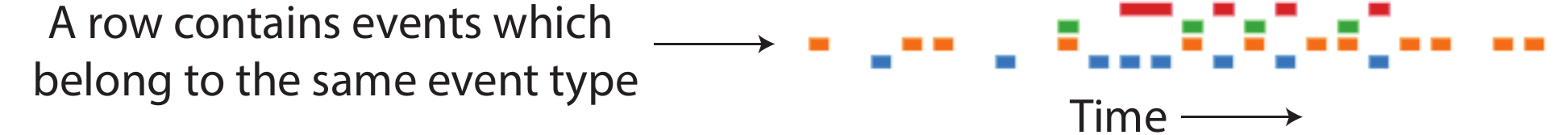}
  \vspace{-7mm}
  \caption{A pixel display in the table view.}
  \vspace{-3mm}
  \label{pixel}
\end{figure}

After the analyst creates a new event type, two major updates occur in Segue's interface: the pixel displays in the table view (Fig.~\ref{teaser}g), and the spatial layout in the network view (Fig.~\ref{teaser}a). In the table view, the events extracted from a dynamic ego-network are visualized as a pixel display in the row where the node is the focal node. In a pixel display (Fig.~\ref{pixel}), the x-axis is time. Each row in the pixel display corresponds to an event type and each pixel in a row represents an event of that type. The width of a pixel encodes the time span of the event. When the analyst creates a new event type, the distance matrix is updated. Segue updates the spatial layout in the network view to reflect the new distance matrix.

\subsubsection{Interpreting the Spatial Layout}

\begin{figure}[h!]
  \centering
  \vspace{-1mm}
  \includegraphics[width=\linewidth]{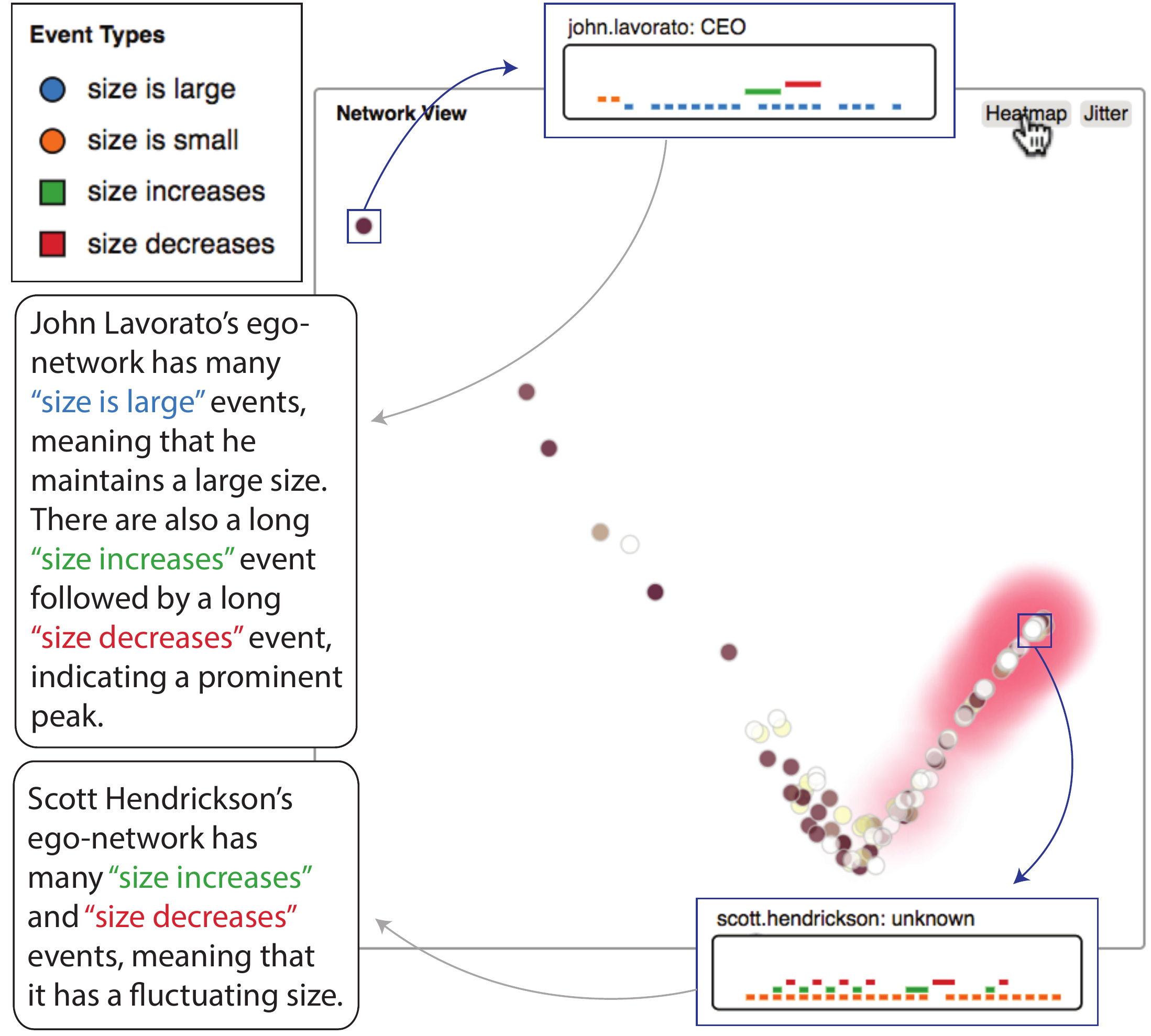}
  \vspace{-7mm}
  \caption{The spatial layout of dynamic ego-networks produced after creating the four event types.}
  \vspace{-1mm}
  \label{oneSpatial}
\end{figure}

After creating all the event types, the analyst examines the spatial layout in the network view (Fig.~\ref{oneSpatial}). As some dots overlap, she clicks on \raisebox{-0.3\height}{\includegraphics[scale = 0.6]{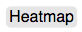}} to overlay a heatmap on the spatial layout to identify areas with overlapping points. The heatmap is rendered using heatmap.js \footnote{{\small \url{https://www.patrick-wied.at/static/heatmapjs/}{}}}. Redder area indicates that more points around the area share the exact same coordinates. Alternatively, users can click on \raisebox{-0.3\height}{\includegraphics[scale = 0.6]{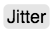}} to jitter the points to reduce visual clutter.

As the analyst hovers over a dot, Segue visualizes the event sequence of the dynamic ego-network as a pixel display. The pixel display reveals the evolution patterns of the ego-network directly in the spatial layout (C3). The analyst observes that there is an isolated dot in the top left, indicating that it is an outlier. When she hovers over this outlier, she finds that it is the dynamic ego-network of John Lavorato, the CEO. His ego-network \raisebox{-0.3\height}{\includegraphics[scale = 0.45]{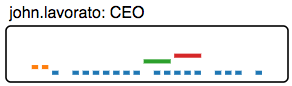}} maintains a large size (there are many blue pixels) and it has one prominent peak (there is a long green pixel followed by a long red pixel). She verifies these findings by clicking on the dot to visualize the timeline visualization of John Lavorato's ego-network in the ego-network view. The timeline visualization does show that there is a prominent peak around May 01 (Fig.~\ref{teaser}b top). She also observes some dynamic ego-networks that have many \raisebox{-0.2\height}{\includegraphics[scale = 0.3]{5_2_sizeissmall}}, \raisebox{-0.3\height}{\includegraphics[scale = 0.3]{5_2_sizeincreases}} and \raisebox{-0.2\height}{\includegraphics[scale = 0.3]{5_2_sizedecreases}} events (e.g., \raisebox{-0.2\height}{\includegraphics[scale = 0.23]{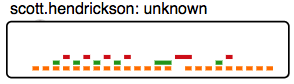}}). These ego-networks are small and fluctuate a lot in size.

\begin{figure}[h!]
  \centering
  \includegraphics[width=0.9\linewidth]{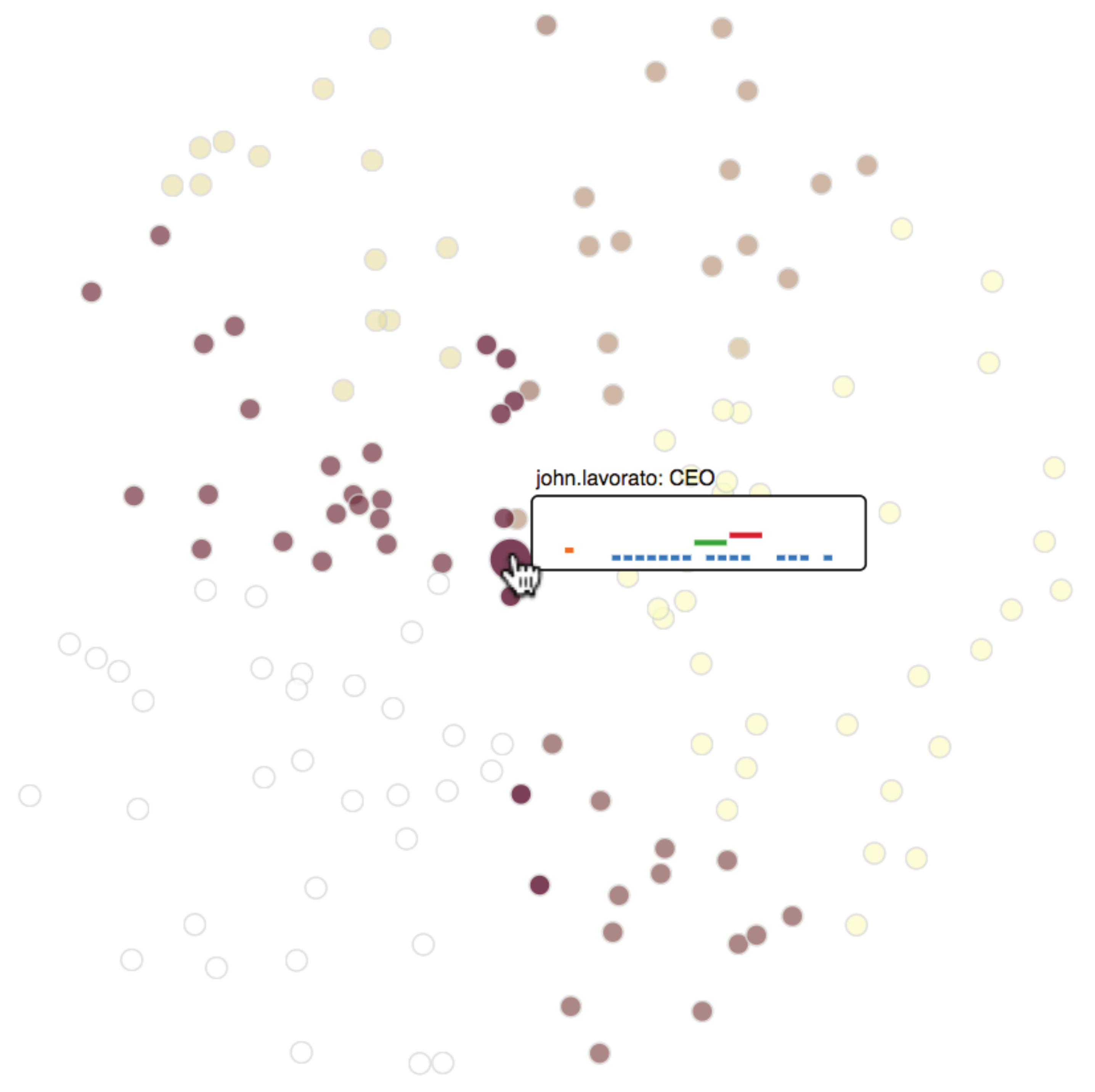}
  \vspace{-3mm}
  \caption{The radial layout produced after double-clicking on the dot that represents John Lavorato's dynamic ego-network.}
  \vspace{-2mm}
  \label{radial}
\end{figure}

The spatial layout is designed with \textit{interpretability} in mind: as analysts hover a point, the event sequence of the dynamic ego-network is visualized. This helps analysts gain insights into why two points are close and why a point is an outlier. Segue also allows analyst to double-click on a node in the spatial layout to convert it to a radial layout (Fig.~\ref{radial}). In the radial layout, the selected node becomes the center node and the distance between another node and the center node encodes the distance derived from their event sequences. Distances in an MDS plot are distorted \cite{interpret1} and we aim to alleviate misinterpretation due to the distortion by allowing analysts to see the undistorted distances in the radial layout.

\section{Usage Scenarios}

\begin{figure*}[t]
  \centering
  \includegraphics[width=\linewidth]{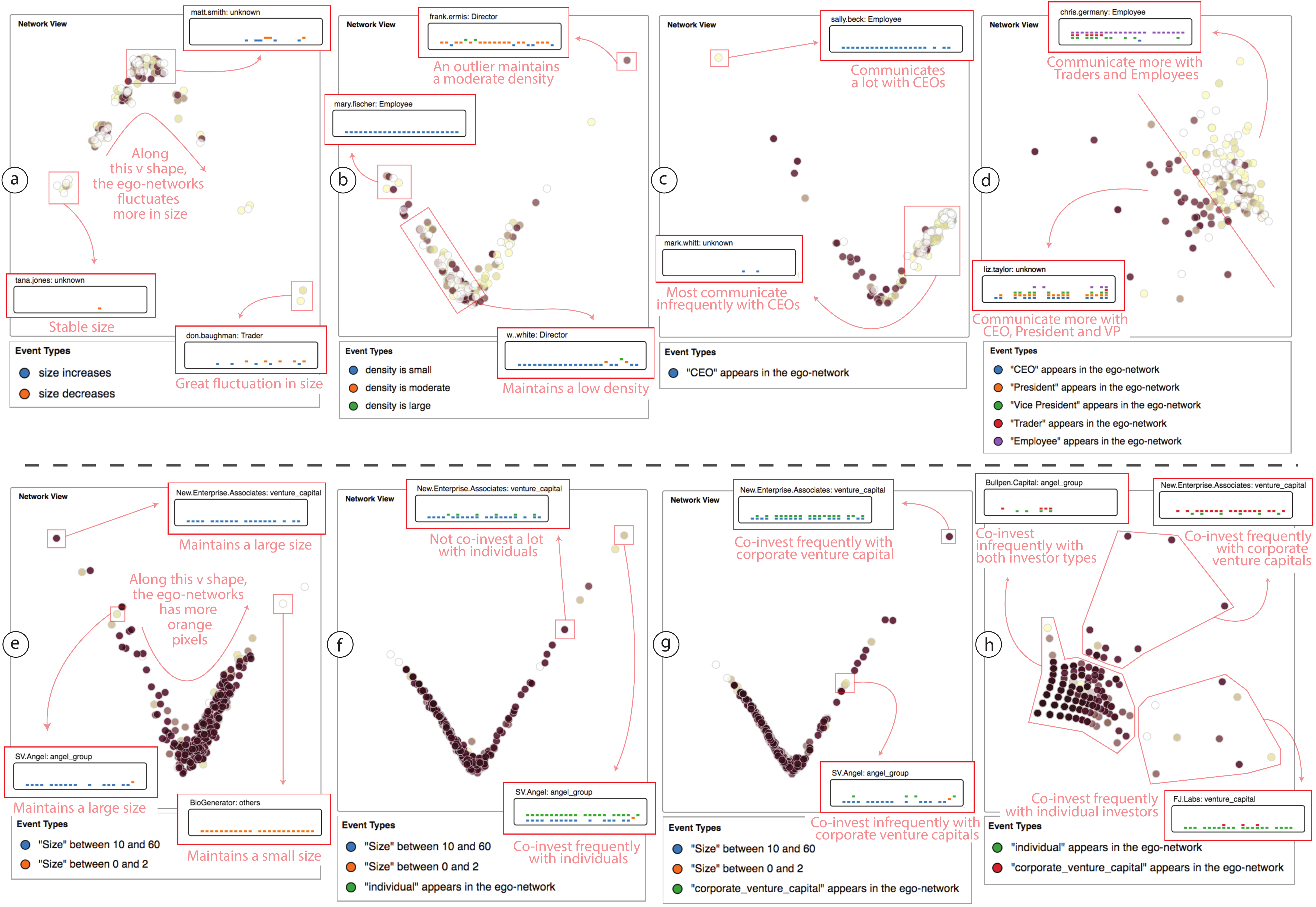}
  \vspace{-7mm}
  \caption{The spatial layouts of dynamic ego-networks created while we are exploring (top) the Enron email network and (bottom) the co-investment network. The red window shows a typical event sequence pattern of the corresponding group of ego-networks.}
  \vspace{-4mm}
  \label{manySpatial}
\end{figure*}

To demonstrate Segue's utility, we provide two usage scenarios. Figure~\ref{manySpatial} shows some of the spatial layouts Segue generates in the two use cases. For a demonstration of the usage scenarios, readers are referred to the videos in the supplemental materials.

\subsection{Enron Email Network}

As mentioned in Section 3.1, there are 142 employees in the dataset, each has a dynamic ego-network. In the interface, we use darker colors to encode upper-level employees (e.g., \raisebox{-0.2\height}{\includegraphics[scale = 0.3]{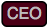}}) and paler colors to encode low-level employees (e.g., \raisebox{-0.2\height}{\includegraphics[scale = 0.3]{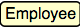}}). Note that a node with the attribute {\small \textbf{\texttt{Employee}}} means the corresponding person is a \textit{low-level} employee in Enron.

We first select the four CEOs from the table to visualize their timeline visualizations in the ego-network view. We observe that their dynamic ego-networks are drastically different and it is very hard to generalize what is common about their evolution patterns. To get a better sense of the data, we try to understand the density and size of the ego-networks.

We want to create a spatial layout where ego-networks with similar fluctuation in size are closer. We construct two event types: \raisebox{-0.3\height}{\includegraphics[scale = 0.6]{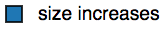}} and \raisebox{-0.3\height}{\includegraphics[scale = 0.6]{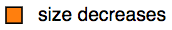}}. To create \raisebox{-0.3\height}{\includegraphics[scale = 0.6]{7_1_sizeincreases}}, we select the size time series and specify the range, slope of size $\geqslant$ 0.36 using the slope slider. As we adjust the slope range, the ego-network view provides a preview of the events that will be extracted if we add the event type. We adjust the slope range until we get a satisfactory result. We click on \raisebox{-0.3\height}{\includegraphics[scale = 0.25]{5_2_1_add}} to create \raisebox{-0.3\height}{\includegraphics[scale = 0.6]{7_1_sizeincreases}}. We then specify slope of size $<$ -0.36 to create \raisebox{-0.3\height}{\includegraphics[scale = 0.6]{7_1_sizedecreases}}. By observing the resulting spatial layout (Fig.~\ref{manySpatial}a), we find that most people have a stable size and there are only a few outliers whose size is highly fluctuating. We select Tana Jones's ego-network whose size is not fluctuating and Don Baughman's ego-network whose size is highly fluctuating to inspect their size time series in the ego-network view. Looking at their size time series, we verify that Don Baughman's size time series does have more ups and downs than Tana Jones's.

To dive deeper into the density distribution of the ego-networks, we create a spatial layout where ego-networks that maintains a comparable density are closers. We construct three event types: \raisebox{-0.3\height}{\includegraphics[scale = 0.6]{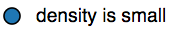}}, \raisebox{-0.3\height}{\includegraphics[scale = 0.6]{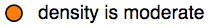}} and \raisebox{-0.3\height}{\includegraphics[scale = 0.6]{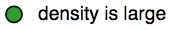}}. They are created by choosing the density time series and specifying 1 $>$ density $\geqslant$ 0, 2 $>$ density $\geqslant$ 1 and 3 $\geqslant$ density $\geqslant$ 2 respectively. From the spatial layout (Fig.~\ref{manySpatial}b), we observe that Frank Ermis maintains a moderate density over time which makes it a clear outlier. We also find that most ego-networks maintain a low density over time.

After getting a brief idea about our data, we delve into more complicated evolution patterns. We are interested in knowing the communication patterns in Enron: who are those who communicate a lot with the executives (e.g., CEOs, presidents and vice presidents) and who are those who communicate a lot with the low-level employees (e.g., traders and employees)? We first select the {\small \textbf{\texttt{CEO}}} time series and create \raisebox{-0.3\height}{\includegraphics[scale = 0.6]{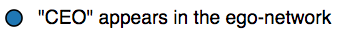}} using the shortcut \raisebox{-0.3\height}{\includegraphics[scale = 0.6]{5_2_1_s}}. The outlier Sally Beck immediately stands out (Fig.~\ref{manySpatial}c). She is only a low-level employee but she communicates very frequently with the CEOs. We wonder why it is the case. We continue to add another event type \raisebox{-0.3\height}{\includegraphics[scale = 0.6]{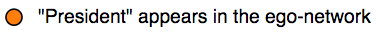}}. While Sally Beck is still an obvious outlier, the CEO John Lavorato becomes another outlier, indicating both Sally and John communicate frequently with the CEOs and presidents. After adding \raisebox{-0.3\height}{\includegraphics[scale = 0.6]{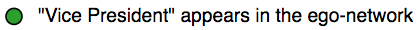}}, Sally Beck moves closer to the other ego-networks. This is because like most other people, Sally does not communicate frequently with vice presidents. Still, Sally Beck, being a low-level employee who communicates frequently with the executes, piques our interest. We searched for her LinkedIn account and discovered that she was indeed a managing director rather than a low-level employee in Enron. We continue to add two more event types \raisebox{-0.3\height}{\includegraphics[scale = 0.6]{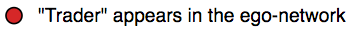}} and \raisebox{-0.3\height}{\includegraphics[scale = 0.6]{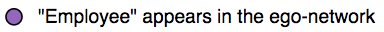}}. The dots seem to separate into two halves (Fig.~\ref{manySpatial}d). One half contains paler dots representing low-level employees. Another half contains darker dots representing upper-level employees. A typical pattern of the half with low-level employees (paler dots) is frequent communication with low-level employees (many have more red or purple pixels) and a typical pattern of the half with upper-level employees (darker dots) is frequent communication with upper-level employees (many have more blue, orange or green pixels).

\subsection{Co-Investment Network}

A start-up needs to obtain funding from investors to grow, survive, and commercialize their ideas. For risk mitigation reasons, there are often multiple investors who co-invest in the same start-up. From crunchbase.com, we collected the co-investment activities among 335 most active US investors from Aug 2014 to Jul 2016 (24 months). Hence, there are in total 335 ego-networks, each with 24 snapshots. In investor A's ego-network in month $T$, investor B is present if A and B co-invested in the same company in that month. An edge in an ego-network snapshot represents a co-investment. Each node has an attribute which represents the investor type. The attribute can be one of {\small \textbf{\texttt{Venture Capital}}}, {\small \textbf{\texttt{Micro Venture Capital}}}, {\small \textbf{\texttt{Individual}}}, {\small \textbf{\texttt{Corporate Venture Capital}}}, {\small \textbf{\texttt{Accelerator}}}, {\small \textbf{\texttt{Angel Group}}}, {\small \textbf{\texttt{Investment Bank}}} and {\small \textbf{\texttt{Others}}}. We use dark brown to encode node attribute which appears the most frequently (i.e. \raisebox{-0.2\height}{\includegraphics[scale = 0.3]{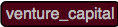}}) and pale yellow to encode node attribute which appears the least frequently (i.e. \raisebox{-0.2\height}{\includegraphics[scale = 0.3]{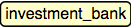}}).

After the dataset is loaded into Segue, we observe immediately from the network view that there is a large number of {\small \textbf{\texttt{Venture Capital}}} nodes. Indeed, 225 out of the 335 US investors are venture capitals. To get a sense of the size distribution of ego-networks, we construct a spatial layout in which the ego-networks that maintain a comparable size are closer. We create two event types \raisebox{-0.3\height}{\includegraphics[scale = 0.6]{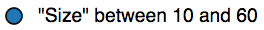}} (meaning large size) and \raisebox{-0.3\height}{\includegraphics[scale = 0.6]{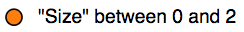}} (meaning small size) by selecting the size time series and specifying two ranges of values. From the projection (Fig.~\ref{manySpatial}e), we observe that there are a few companies which maintain a large size over time (e.g., New Enterprise Associates and SV Angel) and a few companies which maintain a small size (e.g., BioGenerator). We click on some of the large ego-networks to inspect their timeline visualization in the ego-network view. We see that the large investors co-invested with a lot of venture capitals. It is natural because the venture capitals constitute the majority of the 335 US investors.

After the initial exploration, we determine that co-investment with venture capitals is not interesting and we want to investigate how the investors co-invest with other types of investors. We first create \raisebox{-0.3\height}{\includegraphics[scale = 0.6]{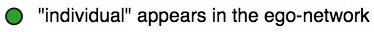}}. This pulls the investors who co-invest with individual investors with comparable frequency closer in the spatial layout (Fig.~\ref{manySpatial}f).

We observe that New Enterprise Associates is now closer to the other investors while SV Angels is still far away from the other investors. This is because like most other investors, New Enterprise Associates does not co-invest frequently with individual investors and unlike most other investors, SV Angels co-invests very frequently with individual investors. We double-click on \raisebox{-0.3\height}{\includegraphics[scale = 0.6]{7_2_individuala}} to remove it and add \raisebox{-0.3\height}{\includegraphics[scale = 0.6]{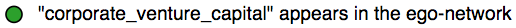}}. This time, SV Angels moves closer to the other investors while New Enterprise Associates moves further away from other investors (Fig.~\ref{manySpatial}g). This is because like most investors, SV Angels does not co-invest a lot with corporate venture capitals and unlike most investors, New Enterprise Associates co-invest frequently with corporate venture capitals.

At this point, we suspect that there might be some strategic differences among the investors regarding the types of investors they co-invest with. We hypothesize that (1) there are investors who co-invest a lot with corporate venture capitals but less with individual investors and (2) there are investors who co-invest a lot with individual investors and less with corporate venture capitals. To verify these hypotheses, we keep only two event types in the event type editor: \raisebox{-0.3\height}{\includegraphics[scale = 0.6]{7_2_individuala}} and \raisebox{-0.3\height}{\includegraphics[scale = 0.6]{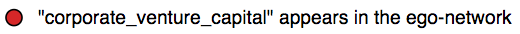}}. Doing so, we create a very interesting spatial layout (Fig.~\ref{manySpatial}h). We identify three different evolution patterns from the plot. The top-right dots consist of investors who co-invest frequently with corporate venture capitals while the bottom-right dots consists of investor who co-invests frequently with individual investors. There are very few dots in the top-right and bottom-right region of the plot, meaning that the two patterns in our hypotheses are rare. Most of the investors are in the left region of the plot. They co-invest infrequently with corporate venture capital and individual investors.

\section{Limitations and Future Work}\label{limitation}

An obvious limitation of our work is a lack of a systematic investigation into the approaches for projecting dynamic ego-networks onto a 2D space. One approach for creating a spatial layout of dynamic ego-networks is demonstrated by egoSlider \cite{egoslider}. It pre-computes a feature vector for each dynamic ego-network based on various network metrics and represents the distances between dynamic ego-networks based on the Canberra distance between the feature vectors. Another approach is to compute the distances between dynamic ego-networks directly from their time series and allow analysts to select the time series type that should be included in distance calculation. A formal study is needed in future to identify the other techniques and understand the trade-offs involved in using different techniques for overviewing evolution patterns of dynamic ego-networks.

Despite a lack of a formal comparative study, our use cases hint on several benefits offered by our data transformation pipeline. First, event sequences produced as a product of the data transformation pipeline enhance \textit{interpretability}. When analysts hover over a dot in the spatial layout, the event sequence visualization provides a simple visual depiction of evolution patterns to help analysts interpret why points are close and why a certain point is an outlier. Other techniques are potentially less interpretable as data structures such as event sequences that can be rendered as simple visuals are not generated as a by-product. Second, compared with the approach that directly computes distances of dynamic ego-networks from their time series, our data transformation pipeline allows analysts to unpack time series into dimensions that are relevant to the analysis. With Segue, analysts can include only the aspects of time series they want to investigate into the distance calculation. For instance, analysts who are interested only in magnitude of the size time series can create the event types ``size is large'' and ``size is small''. If analysts are interested only in the fluctuation of the size time series, they add the event types ``size increases'' and ``size decreases''. Finally, creating event types provides flexibility for analysts to incorporate their domain knowledge into spatial layouts. For example, analysts can specify, based on their domain knowledge, how large an ego-network is large when defining the event type ``size is large''. A comparative study of our data transformation pipeline and other techniques will be required to verify these benefits. 

These potential benefits come with clear sacrifices, the first being information loss. Our data transformation pipeline trades precision in the distance calculation for interpretability. The distance calculation becomes less precise due to information loss in transforming time series into event sequences. To mitigate this problem, we provide the ego-network view for viewing the details of each dynamic ego-network. Developing techniques to produce spatial layouts that are interpretable and are precise in depicting distances between objects will be an interesting future work. Furthermore, although our technique helps analysts see what they want to see by allowing them to include only the relevant aspects of dynamic ego-networks into the distance calculation, it is less suitable for analysts who have a vague idea of what to analyze. Automatic approaches to creating spatial layouts (e.g., computing distances based on pre-computed features of dynamic ego-newtorks as in egoSlider \cite{egoslider}) are more suitable when users do not know what spatial layouts to create.

Our techniques compute the distances between event sequences by converting them into feature vectors and computing the Euclidean distance between them. While prior work \cite{eventaction} and the two usage scenarios show that it is effective for revealing interesting global patterns in a large number dynamic ego-networks, we envision several ways to increase the utility of Segue by enhancing the distance function beyond edit distance. One possible extension is to allow analysts to set different weights for each feature so that the evolution pattern that corresponds to a feature with a higher weight is more salient in the spatial layout. Also, as noted by Du et al. \cite{peerfinder}, different domains require different similarity measures to address their specific problems. To accomodate for the needs of different users, another future work is to incorporate more similarity measures such as Jaccard index \cite{leoflow} for computing the distances between the extracted event sequences.

A limitation of using Euclidean distance for computing the distance between dynamic ego-networks concerns cluttered points in the resulting embedding of dynamic ego-networks. When only one to two event types are created by analysts, it is likely that many ego-networks share the same feature vector. This impairs analysis as many dots have a distance of zero and hence overlap at the same coordinates. As analysts add more event types, visual clutter will be reduced as the feature vectors of dynamic ego-networks will become more different. There are two major approaches to handle overlapping points when there are few event types: \textit{advanced distance functions} and \textit{visual augmentation} of the spatial layout. Advanced distance functions can potentially reduce visual clutter even when the number of event types is small as they provide better differentiating power of event sequences. While Euclidean distance seems to be sufficient in our usage scenarios, it considers two dynamic ego-networks the same even they have very different event sequences. For example, the event sequence {\small \textbf{\texttt{[A, B, A]}}} where {\small \textbf{\texttt{A}}} and {\small \textbf{\texttt{B}}} are events and the duration between two consecutive events is seven days, and {\small \textbf{\texttt{[A, B, A]}}} where the duration between two consecutive events is 3 months have a Euclidean distance of zero. Advanced distance functions such as M\&M measure \cite{mAndM} would consider these two event sequences different by capturing their differences in durations between consecutive events. Another way to reduce the impact of visual clutter is to use visual augmentation. Currently, analysts can overlay a heatmap on the spatial layout to identify areas where there are many overlapping points. They can also jitter the points to reduce visual clutter. Other techniques include automatically annotating and applying visual cues to the spatial layout in which serious visual clutter appears.

While we have demonstrated how Segue enables analysts to construct spatial layouts for a dataset with around 300 dynamic ego-networks, we have yet to discuss the scalability of this technique. A large number of ego-networks or a large number of node attributes will render Segue ineffective. When there is a large number of ego-networks, it requires a long time for classical MDS to project the dynamic ego-networks onto a low-dimensional space. In our trials, the projection can be done interactively without much latency when there are around 500 dynamic ego-networks. When there is a large number of node attributes, an obvious challenge is the limited number of color channels. Segue uses color to distinguish among both event types and node attributes. When there are many node attributes, there will be many different colors in Segue's interface. Users may find the relationship of which color encodes what confusing. Further user studies would help identify other outstanding usability issues.

\section{Conclusion}

We have presented Segue, a visual analysis system that enables analysts to \textit{interactively} construct \textit{interpretable} spatial layouts of dynamic ego-networks. Through two usage scenarios, we have demonstrated how Segue empowers analysts to get an overview of the evolution patterns of many ego-networks by creating various spatial layouts. Making sense of dynamic networks poses a constellation of challenges to the visualization community. We hope that Segue inspires new techniques and follow-on studies aimed at reducing the roadblocks to analyzing new types of dynamic networks.

\bibliographystyle{abbrv-doi}

\bibliography{template}
\end{document}